# SOFTWARE CLONING IN EXTREME PROGRAMMING ENVIRONMENT

Ginika Mahajan*

Ashima**

## ABSTRACT

*Software systems are evolving by adding new functions and modifying existing functions over time. Through the evolution, the structure of software is becoming more complex and so the understandability and maintainability of software systems is deteriorating day by day. These are not only important but one of the most expensive activities in software development. Refactoring has often been applied to the software to improve them. One of the targets of refactoring is to limit Code Cloning because it hinders software maintenance and affects its quality. And in order to cope with the constant changes, refactoring is seen as an essential component of Extreme Programming. Agile Methods use refactoring as important key practice and are first choice for developing clone-free code. This paper summarizes my overview talk on software cloning analysis. It first discusses the notion of code cloning, types of clones, reasons, its consequences and analysis. It highlights Code Cloning in Extreme Programming Environment and finds Clone Detection as effective tool for Refactoring.*

*Computer Science and Engineering Department, Thapar University, Patiala.

**Assistant Professor, Computer Science and Engineering Department, Thapar University, Patiala.

# 1. INTRODUCTION

Software Reuse is defined as the "the process of creating software systems from existing software systems "[9]. This notion is shared worldwide that it's easier to modify the existing software than developing programs from the scratch and so reuse is emerging as a central theme within Software Engineering. The use of existing components is done basically with the activity of copy and paste.

The practice of copy and paste code is extensively acknowledged but is rarely explicitly accounted for in models of software development. Copy and paste programming is a common activity but it introduce a negative point to reuse by creating clones. Cloning, the copying and modifying of blocks of code, is the basic means of software reuse. Various programming methodologies are used in software development process. As Extreme Programming methods are primarily used to decrease the time needed to bring software products to market, so Extreme Programming methodology is accompanied with high degree of reuse. Refactoring is one of the main practices of Extreme Programming and thus refactoring is used in the cloning process.

Code clones are considered harmful in software development, also provides hindrance in software evolution and maintenance phase. The predominant approach is to try to eliminate them through refactoring.

## 1.1 Software Reuse

Software reuse reduces software development and maintenance costs in the process of creating software systems. It is the likelihood a segment of source code that can be used again to add new functionalities with slight or no modification. Reusable modules and classes reduce implementation time, increase the likelihood that prior testing and use has eliminated bugs and localizes code modifications when a change in implementation is required. Basically, the reuse of a software artifact is its integration into another context. The purpose of reuse is to reduce cost, time, effort, and risk; and to increase productivity, quality, performance, and interoperability [18]. One form of reuse is to copy-paste the code which results in duplication of code i.e. clones

## 1.2 Software Cloning

The copying of code has been studied within software engineering mostly in the area of clone analysis. Software clones are regions of source code which are highly similar; these regions of similarity are called clones, clone classes, or clone pairs. While there are several reasons why two regions of code may be similar, the majority of the clone analysis literature attributes cloning activity to the intentional copying and duplication of code by programmers [5]; clones may also be attributable to automatically generated code, or the constraints imposed by the use of a particular framework or library [30]. In addition to these, some other issues, including programmers' behavior such as laziness and the tendency to repeat common solutions, technology limitations, code understandability and external business forces have influences on code cloning [26].

Cloning works at the cost of increasing lines of code without adding to overall productivity. Same software bugs and defects are replicated that reoccurs throughout the software at its evolving as well its maintenance phase. It results to excessive maintenance costs as well. So cut paste programming form of software reuse deceivingly raise the number of lines of code without expected reduction in maintenance costs associated with other forms of reuse. So, to refactor code clones, is a promising way to reduce the maintenance cost in future.

### 1.3 Refactoring

Refactoring has as many definitions as practitioners, but perhaps the most concise and certainly the most widely cited definition is as follows: "A change made to the internal structure of software to make it easier to understand and cheaper to modify without changing its observable behavior." [26]. A more practical-minded definition is, "a technique in which a software engineer applies well-defined source-level transformations with the goal of improving the code's structure and thus reducing subsequent costs of software evolution." [34].

The concept of refactoring (and also the word "refactoring" itself) was coined already several years ago, but its breakthrough came with the integration of refactoring into the software development process Extreme Programming [21].

### 1.4 Extreme Programming

The use of a series of small and systematic transformations during the software development does not fit into the waterfall or the spiral models of software engineering [3]. Instead it corresponds to a short-cycled, iterative method. It is therefore no coincidence that refactoring became prominent together with Extreme Programming, where development cycles are kept as short as possible.

Extreme Programming (XP) is a software development methodology which interleaves design and coding, rather than considering them as serial processes. Proponents of XP have argued that much time in traditional software development is wasted developing extensible designs which are never utilized. Furthermore, in traditional software development there is no established route for changes to requirements to propagate through specifications and design to code [13].

## 2. BASIC CONCEPT OF CLONE DETECTION

Clones, as the name implies, are copied regions of code. However, unlike a biological clone, a software clone may or may not be exactly the same [25]. For a given clone relation, a pair of code portions is called clone pair if the clone relation holds between the portions. An equivalence class of clone relation is called clone class [34]. The code clone pair and class are shown in fig.1

### 2.1 Code Fragment and Code Clone

**Definition 1:** Code Fragment. A code fragment (CF) is any sequence of code lines (with or without comments). It can be of any granularity, e.g., function definition, begin end block, or sequence of statements [4].

**Definition 2:** Code Clone. A code fragment CF2 is a clone of another code fragment CF1 if they are similar by some given definition of similarity, that is, f (CF1) = f (CF2) where f is the similarity function [4].

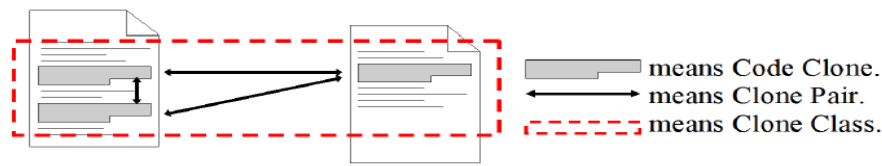

**Fig.1 Clone relationship [4]**

### 2.2 Types of Clone

The degree of similarity varies from an exact superficial copy to more semantically similar regions of code (they do the same thing) or structurally similar regions of code (similar patterns of statements) [25]. Clones are categorized into four types

**Type I: Exact Software Clones (Changes in layout and formatting)**

In Type I clone, a copied code fragment is the same as the original. However, there might be some variations in whitespace (blanks, new line(s), tabs etc.), comments and/or layouts.

**Type II: Near-Miss Software Clone (Renaming Identifiers and Literal Values)**

A Type II clone is a code fragment that is the same as the original except for some possible variations about the corresponding names of user-defined identifiers (name of variables, constants, class, methods and so on), types, layout and comments.

**Type III: Near-Miss Software Clone (Statements added/deleted/modified in copied fragments)**

Type III is copy with further modifications, e.g. a new statement can be added, or some statements can be removed. The structure of code fragment may be changed and they may even look or behave slight differently.

**Type IV: Near-Miss Software Clone (Statements reordering/control replacements)**

Type IV clones are the results of semantic similarity between two or more code fragments. In this type of clones, the cloned fragment is not necessarily copied from the original. The same kind of logic may be implemented making the code fragments similar in their functionality [5]. Example study of Clone types shown in Table 1

**2.3 Software Clone Analysis**

Clone analysis studies report a number of values indicating how much code was cloned. In their extensive review of clone analysis research, Roy and Cordy (2007) report that between 5% and 20% of code in large-scale software systems is duplicated or copied [5][25]. Baker reported up to 38% of code in the systems she studied was cloned [2]; Ducasse et al. (1999) found 59.3% of the system they studied was copied code. Clone analysis has typically been used to measure the quality of source code, and indicate areas for refactoring [1] and improvement. The general argument against cloning (i.e., the duplication of similar or identical code) is that similar regions of code represent unnecessary duplication within the program, which increases the cost and complexity of maintaining the code as programmers must track more code, make repeated edits to multiple clones, and can easily propagate bugs [25].

**Example code fragment**

```
void sumProd(int n) {            //s0
    int sum=0;                   //s1
    int product=1;               //s2
    for (int i=1; i<=n; i++) {   //s3
        sum=sum + i;             //s4
        product = product * i;   //s5
        fun(sum, product); }}    //s6
```

**Type 1:**

```
void sumProd(int n) {            //s0
    int sum=0;                   //s1
    int product =1;              //s2
    for (int i=1; i<=n; i++) {   //s3
        sum=sum + i;             //s4
        product = product * i;   //s5
        fun(sum, product); }}    //s6
```

```
void sumProd(int n) {            //s0
    int sum=0;                   //s1
    int product =1;              //s2
    for (int i=1; i<=n; i++)     //s3
        sum=sum + i;             //s4
        product = product * i;   //s5
        fun(sum, product); }}    //s6
```

```
void sumProd(int n) {            //s0
    int sum=0;                   //s1
    int product =1;              //s2
    for (int i=1; i<=n; i++) {   //s3
        sum=sum + i;             //s4
        product = product * i;   //s5
        fun(sum, product); }}    //s6
```

           **Changes in comments**      **Changes in formatting**

**Type 2:**

```
void addTimes(int n) {           //s0
int add=0;                       //s1
int times =1;                    //s2
for (int i=1; i<=n; i++) {       //s3
    add=add + i;                 //s4
    times = times * i;           //s5
    fun(add, times); }}          //s6
```

```
void sumProd(int n) {            //s0
double sum=0.0;                  //s1
double product =1.0;             //s2
for (int i=1; i<=n; i++) {       //s3
    sum=sum + i;                 //s4
    product = product * i;       //s5
    fun(sum, product); }}        //s6
```

   **Renaming of identifiers**   **Renaming of Literals and Types**

**Type 3:**

```
void sumProd(int n) {            //s0
int sum=0;                       //s1
int product =1;                  //s2
for (int i=1; i<=n; i++) {       //s3
    if (i % 2 == 0) sum+= i;     //s4m
    product = product * i;       //s5
    fun(sum, product); }}        //s6
```

```
void sumProd(int n) {            //s0
int sum=0;                       //s1
int product =1;                  //s2
for (int i=1; i<=n; i++)         //s3
    if (i % 2 == 0){             //s3b
    sum=sum + i;                 //s4
    product = product * i;       //s5
    fun(sum, product); }}        //s6
```

```
void sumProd(int n) {            //s0
int sum=0;                       //s1
int product =1;                  //s2
for (int i=1; i<=n; i++) {       //s3
    sum=sum + i;                 //s4
    //s5  line deleted
    fun(sum, product); }}        //s6
```

   **Modification of lines**   **Addition of new of lines**   **Deletion of lines**

**Type 4:**

```
void sumProd(int n) {            //s0
int sum=0;                       //s1
int product =1;                  //s2
int i = 0;                       //s7
while (i<=n) {                   //s3'
    sum=sum + i;                 //s4
    product = product * i;       //s5
    fun(sum, product);           //s6
    i =i + 1; }}                 //s8
```

```
void sumProd(int n) {            //s0
int product =1;                  //s2
int sum=0;                       //s1
for (int i=1; i<=n; i++) {       //s3
    sum=sum + i;                 //s4
    product = product * i;       //s5
    fun(sum, product); }}        //s6
```

**Table 1: Types of Clone**

### 2.4 Overview of Clone Detection Tools and Techniques

Code cloning detection had been an active research for almost two decades. The detection of code clones is a two phase process which consists of a transformation and a comparison phase. In the first phase, the source text is transformed into an internal format which allows the use of a more efficient comparison algorithm. During the succeeding comparison phase the actual matches are detected. Due to its central role, it is reasonable to classify detection techniques according to their internal format.

This section gives an overview of the different techniques available for each category while selecting a representative for each category [13][16][19][20][22].The selected techniques cover the whole spectrum of the state of the art in clone detection. The techniques work on text, lexical and syntactic information, software metrics, and program dependency graphs [4].

1. **String-based**- the program is divided into a number of strings (typically lines) and these strings are compared against each other to find sequences of duplicated strings.
2. **Token-based**- a laxer tool divides the program into a stream of tokens and then searches for series of similar tokens.
3. **Parse–tree based**- after building a complete parse-tree one performs pattern matching on the tree to search for similar sub–trees.
4. **PDG based**- after obtaining program dependency graph similar graphs are search.
5. **Metric based** -metrics are calculated form program and these are used to find duplicate code.
6. **Hybrid**- detection techniques that use a combination of the other clone detection techniques.

## 3. CLONING IN EXTREME PROGRAMMING ENVIRONMENT

The pace of change in the software development industry remains at a high rate. People continue to push the boundaries of known techniques and practices in an effort to develop software as efficiently and effectively as possible. Extreme Programming and Agile Software Methodologies have emerged as an alternative to comprehensive methods designed primarily for very large projects. Teams using XP are delivering software often and with very low defect rates [24].

### 3.1 Extreme Programming and Reuse

XP does not address explicitly the issue of software reuse as one of its practices. This may wonder since many believe that software reuse provides "the key to enormous savings and benefits in software development" [31]. XP per se does not aim at developing software for possible future reuse in order to avoid overhead during development[14].

On the other hand XP – compared to more traditional development methodologies - intrinsically guides software engineers to develop software, which is of high quality and therefore suited for ad-hoc reuse. In particular the practice of continuous refactoring may improve internal quality metrics and affect reusability of a software system in a positive way [31].

### 3.2 Extreme Programming and Refactoring

In Extreme Programming (XP) much emphasis is given on an agile, iterative and customer oriented way of how to develop software. Among the top priorities of XP are (a) customer satisfaction through continuous delivery of valuable software and (b) embracing changing requirements. The practices of XP are tailored to achieve such goals: iterative and informal planning, simple design, continuous refactoring of the code, pair programming, test first and continuous integration – just to mention a few [22]. Most of these practices are intended to be used during development and maintenance where refactoring is one of the important among these.

Continuous refactoring is an element of XP development where developers make continuous improvements to maintain the code. It doesn't change or improve software from a functional point of view; the program is intended to do the same thing after refactoring that it did before refactoring.

### 3.3 Extreme Programming and Cloning

Production software developed with extreme programming practices has fewer exact clones than software developed in more typical ways, as well as fewer near-clones which allow duplication to be removed in a simple, straightforward manner. The refactoring opportunities that persist in XP-developed code often require the introduction of a new, higher-level construct to remove the duplication present. XP test code provides many of the same kinds of refactoring opportunities as non-XP code, but in smaller quantity [36]. One of the targets of refactoring is code clone.

The Extreme Programming (XP) community has integrated frequent refactoring as a part of the development process and has argued that fewer clones are found in XP process software[12] [27] as shown in table 2.

| Project | XP Use | Avg. Score | Avg. # Nodes | In Test Code | Exact clones | Exact Clones In Test Code |
|---|---|---|---|---|---|---|
| Alcatraz | None | 110 | 154 | | 5 | |
| Donquixote | None | 381 | 471 | | 19 | |
| Spandex | None | 143 | 232 | | 2 | |
| Iago | Semi | 165 | 249 | 95% | 0 | 0 |
| Pdq | Semi | 81 | 167 | 76% | 0 | 0 |
| Ontology | Full | 50 | 94 | 33% | 5 | 3 |
| Tantalus | Full | 62 | 110 | 24% | 1 | 0 |
| Ardor | Full | 33 | 56 | 91% | 0 | 0 |
| Ardor_Test | Full | 65 | 102 | N/A | 4 | 4 |

Table 2: Comparison of Clones in XP, semi-XP and non-XP projects [12]

### 3.4 Cloning and Refactoring

Refactoring code clones is an effective way to reduce code clones in a software system [11].Refactoring patterns are used to find the clone codes. It is a typical activity to remove code clones. The patterns get code clones into common routine like method by using distinctive functions.

"Every pattern describes a problem which occurs over and over again in our environment, and then describes the core of the solution to that problem, in such a way that you can use this solution a million times over, without ever doing it the same way twice." – Christopher Alexander

Various refactoring patterns are used to remove code clones like [26] "Extract Method" and "Pull Up Method" that are related to the code clone. However, quite often one must use a series of refactorings to actually remove duplicated code, as in Transform Conditionals into Polymorphism where duplicated conditional logic is refactored over the class hierarchy using polymorphism [8]. With refactoring tools like the refactoring browser [7] emerging from research laboratories into mainstream programming environments, refactoring is becoming a mature and widespread technique.

**3.4.1 Extract Method**: To put it plainly, "Extract Method" means extraction of a part of existing method as a new method, and extracted part is replaced by a new method caller shown in fig 2.

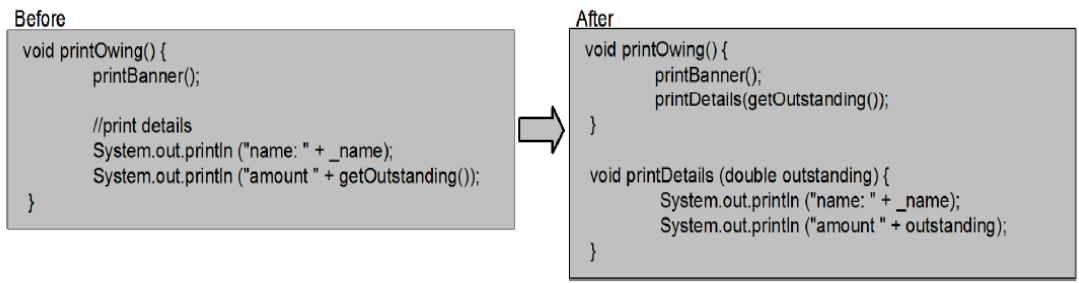

Fig 2. Example of Extract Method [36]

In general, this pattern is applied to the case that there is a too long method. In applying the pattern to code clones, a new method, that is a code fragment of code clone, is defined and the original code clones are replaced by the new method caller. As the result, we can remove the code clones[36].

**3.4.2 Pull Up Method:** "Pull Up Method" is a simple refactoring pattern. It means pulling up a method which defined in child class to its parent class. If the parent class has several child classes and some of them have the same method (that is, code clone), pulling up the method can remove the code clone[36].

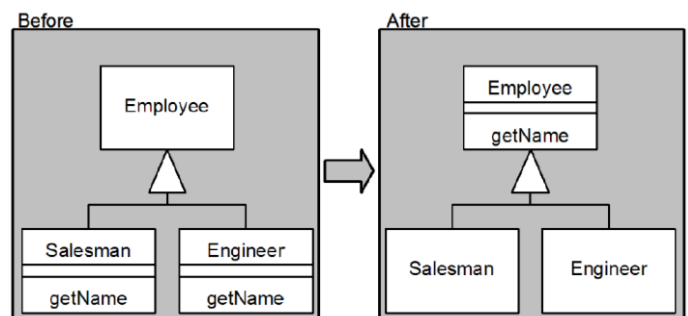

Fig 3. Example of Pull Up Method [36]

**3.5 Cloning and Software Maintenance**

The impact of clones is of special concern from a software maintenance point of view.

The software life cycle comprises two major parts; first we define the specification and implement it; then, we need to maintain the finished product and evolve it to better suit user needs. However for software development it has been found that maintenance and evolution are also critical activities from the cost perspective and might comprise upto 80% of the overall cost and effort [15][17].

Fowler suggests that code duplication or cloning is a bad smell and thus one of the major indicators of poor maintainability [15]. Cloning is an easy, tempting alternative to the hard work of actually refactoring the code. Maintaining the cloning relationship is thus a very important consequence of copying and pasting. Without tracking clones over time, identifying and consistently changing clones can be problematic [28][29] and increases the maintenance cost.

## 4. CONCLUSION

Clone analysis research typically comes in one of two varieties, either research on improving the algorithms in terms of performance and accuracy, or clone analytic studies which discuss the application of clone detection tools on software systems. There has been a broad assumption that code clones are inherently bad and refactoring would remove the problems of clones. Effective refactoring can be easily introduced with software cloning for the removal of maintenance related issues of software.

Software reuse is a key success factor for software development and should be supported as much as possible by the development process itself. Refactoring is very valuable for intrinsically delivering code, which is easier to reuse than code which has not been refactored. Agile Methods already use refactoring as one of their key practices and could be a first choice for developing code in a way that supports, among other benefits such as good maintainability and reusability.

Refactoring tools which identify and operate on simple duplication have already been employed. However, it seems likely that other tools will be needed to assist XP developers to perform more subtle refactoring for clone detection.